\documentclass[slac_one]{revtex4}
\usepackage{graphicx}
\usepackage{fancyhdr}
\newcommand{\numutau}{\nu_{\mu}\rightarrow\nu_{\tau}}
\newcommand{\numunue}{\nu_{\mu}\rightarrow\nu_{e}}
\newcommand{\nutau}{\nu_{\tau}}
\begin{document}

\title{The OPERA Long Baseline Experiment: Status and First Results} 

%

\author{D. Duchesneau (for the OPERA Collaboration)}
\affiliation{LAPP, Universit\'e de Savoie, CNRS/IN2P3, Annecy-le-Vieux, France}

\begin{abstract}
OPERA (Oscillation Project with Emulsion tRacking Apparatus)
is an international collaboration between Europe and Asia, aiming to give the
first direct proof of tau neutrino appearance in a pure muon
neutrino beam, in order to validate the hypothesis for atmospheric neutrino oscillations.
The first european long baseline neutrino beam called  CNGS is produced at CERN 
and sent in the direction of the Gran Sasso underground laboratory 730 km away,
 where the OPERA detector is located.
 Since 2006  the electronic detector part is fully commissioned
and running. Cosmic ray events have been recorded on a regular basis  
and the first  neutrino beam events have been observed in the target elements
made of very precise emulsion films and lead sheets during the last run in autumn 2007. 
This paper reviews the status of the detector, the beam performances,  the first
results from  the neutrino  event analysis and the prospects.
\end{abstract}

\maketitle

\thispagestyle{fancy}

\section{INTRODUCTION} 
OPERA~\cite{OPERA} is a long baseline neutrino experiment located
 in the Gran Sasso underground laboratory (LNGS) in Italy. The collaboration is composed
of about 200 physicists coming from 36 institutions in 13 different countries. 
The experiment is a massive hybrid detector
 with nuclear emulsions used as very precise tracking devices and electronic detectors 
to locate the neutrino interaction events in the emulsions.
 It is designed to primarily search
for $\nu_{\tau}$ appearance in the CERN high energy $\nu_{\mu}$ beam CNGS~\cite{CNGS} at 
730 km from the neutrino source,
in order to establish unambiguously  the origin of  the neutrino
 oscillations observed at the "atmospheric" $\Delta m^{2}$ scale. The preferred hypothesis 
to describe this phenomenon being  $\nu_{\mu} \rightarrow \nu_{\tau}$ oscillation.
Combining all the present known neutrino data 
 the best fit values of a global three flavour 
analysis of neutrino oscillations~\cite{Fogli2008} give for 
 $\nu_{\mu} \rightarrow \nu_{\tau}$ oscillation parameters 
 $\Delta m^{2}=2.39$x$10^{-3}$$ \mathrm{eV}^{2}$  and $\mathrm{sin}^{2}2\theta$=0.995.
The range of allowed values at 3 $\sigma$  is  
2.06x$10^{-3} < \Delta m^{2} <$ 2.81x$10^{-3}$$ \mathrm{eV}^{2}$.
In addition to the dominant $\numutau$ oscillation in $\nu_{\mu}$ beam, it is possible that a 
sub-leading $\numunue$ transition occurs as well.
This process will also be investigated by OPERA profiting from its 
excellent electron identification capabilities 
to asses a possible improvement on the knowledge of the third yet unknown mixing angle $\theta_{13}$.

The $\nutau$ direct appearance search is based on the observation
of events produced by  charged current interaction (CC) with the
$\tau$ decaying in leptonic and hadronic modes.
In order to directly observe the $\tau$ kinematics,
the principle of the OPERA experiment is to observe the $\tau$ trajectories 
and the decay products in emulsion films composed of
two thin emulsion layers (44 $\mu$m thick) put on either side of a plastic base
 (205 $\mu$m thick). The detector concept which is described in the next section combines
micrometer tracking resolution, large target mass together with  good lepton identification. 
This concept allows to reject efficiently the main topological background coming from charm production in 
$\nu_\mu$ charged current interactions.

\section{DETECTOR OVERVIEW}

\begin{figure*}[htb]
\vspace{-0.3cm}
\begin{center}
    \includegraphics[width=17cm]{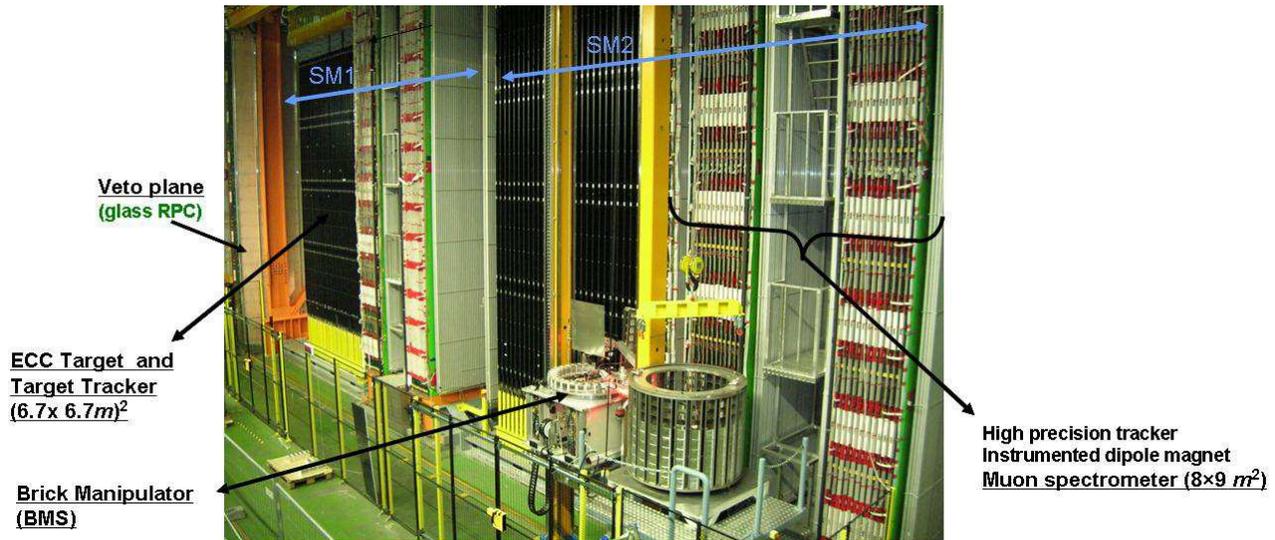}
\end{center}
\vspace{-0.8cm}
\caption{View of the OPERA detector in Hall C of the Gran Sasso Underground Laboratory in May 2007.}
\label{fig:opera}
\vspace{0cm}
\end{figure*}

The OPERA detector is installed in the Hall C of the Gran Sasso underground laboratory.
Figure~\ref{fig:opera} shows a recent picture of the detector which is 20 m long with a 
cross section of about 8x9 $\mathrm{m}^{2}$ and composed 
of two identical parts called super modules (SM). Each SM has a target section and
a muon spectrometer. 

The spectrometer  allows a determination of the charge and momentum
of muons going through by measuring their curvature in  a 
dipolar magnet made of 990 tons of iron, and providing 1.53 Tesla transverse to the 
neutrino beam axis. Each spectrometer is  equipped with six vertical planes of 
drift tubes as precision tracker together with 22 planes (8x8 $\mathrm{m}^{2}$) 
of RPC bakelite chambers reaching a spatial resolution of $\sim$1 cm and an efficiency of 96\%.
The precision tracker planes are composed of 4 staggered layers of 168 aluminium tubes,
8 m long  with 38 mm  outer  diameter. The spatial resolution
of this detector is better than 500 $\mu$m.
The physics performance of the complete spectrometer should reduce the charge confusion 
to less than 0.3\% and gives a momentum resolution better than 20\% for momentum less than
50 GeV. The muon identification efficiency reaches 95\% adding  the target
tracker information for the cases where  the  muons stop inside the target. 

\begin{figure*}[htb]
\vspace{-0.3cm}
\begin{center}
    \includegraphics[width=6cm]{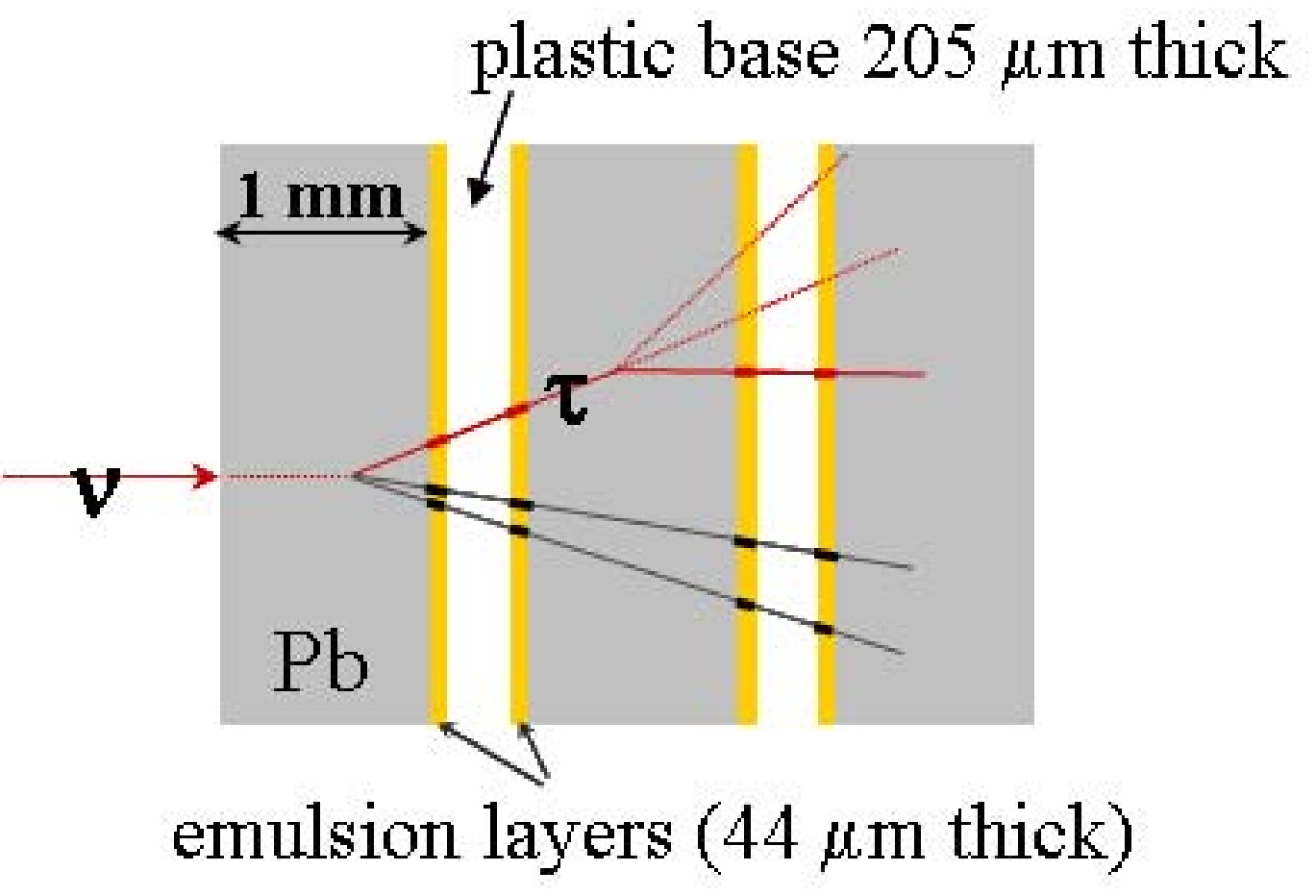}
    \includegraphics[width=6cm]{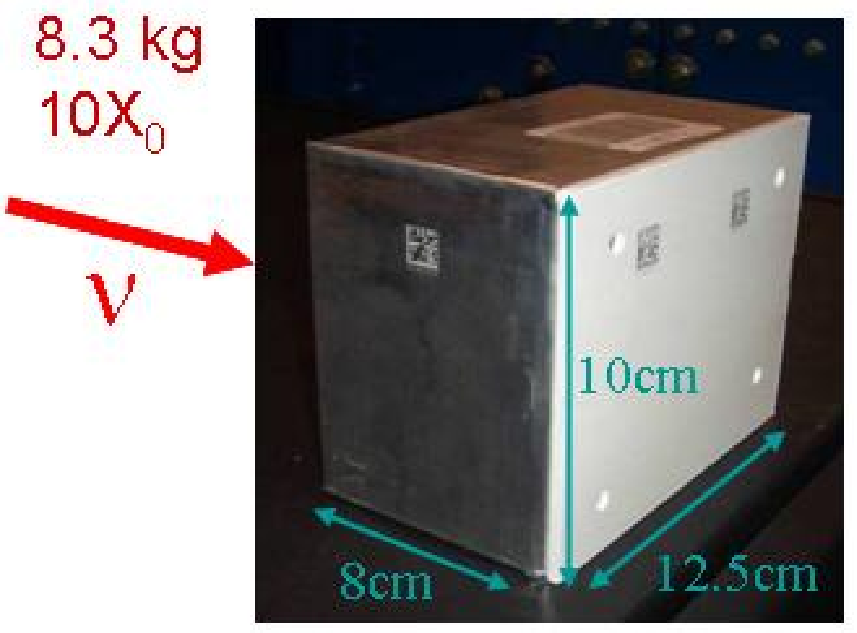}
\end{center}
\vspace{-0.8cm}
\caption{a)Schematic structure of an ECC cell. The $\tau$ decay kink is
reconstructed by using four track segments in the emulsion films. b) Picture of an assembled brick. 
 Each brick weights about 8.6 kg and has a thickness of 10 radiation length $X_{o}$.          }
\label{fig:brick}
\vspace{0cm}
\end{figure*}

The target section is composed of 31 vertical light supporting steel structures, called walls,
interleaved with double layered planes of 
 6.6 m long scintillator strips in the two transverse
directions. The main goals
of this electronic detector are to provide a trigger for the
neutrino interactions, an efficient event pattern recognition together with the magnetic spectrometer
allowing a clear classification of the $\nu$ interactions
 and a precise localisation of the event.
The electronic target tracker spatial resolution
reaches $\sim$0.8 cm and has an efficiency of 99\%.\\
The walls contain the basic target detector units, called ECC brick, sketched in Fig.~\ref{fig:brick}
which are obtained by stacking 56 lead plates with 57 emulsion films. This structure provides
many advantages like a massive target coupled to a very precise tracker, as well as a standalone
detector to measure electromagnetic showers and charged particle momentum using the multiple
coulomb scattering in the lead. The ECC concept has been
already succesfully used for the direct $\nu_\tau$ observation perfomed in 2000 by the DONUT 
experiment~\cite{donut}.

 Behind each brick, an  emulsion film doublet, called Changeable Sheet (CS) is attached in 
 a separate enveloppe. The CS can be detached from the brick for analysis
  to confirm and locate the tracks produced in neutrino interactions.

By the time of this conference, 146500  bricks (1.25 kton of target) assembled
 underground at an average rate of about 700 bricks/day by
a dedicated fully automated Brick Assembly Machine (BAM) with precise robotics  were
installed in the support steel structures from the sides of the walls
using two automated manipulator systems (BMS) running on each side of the experiment. \\
When a candidate brick has been located by the electronic detectors, 
the brick is removed using the BMS and the changeable
sheet is detached and developped. The film is then scanned to 
search for the tracks originating from the neutrino interaction. If none
are found then the brick is left untouched and another
one is removed. When a neutrino 
 event is confirmed the brick is exposed to cosmics
to collect enough alignment tracks before going to the development. 
After development the emulsions are sent to the scanning laboratories hosting
automated optical  microscopes in Europe and Japan, each region using
a different technology~\cite{scan1,scan2}. This step is the 
start of the detailed analysis consisting of finding the neutrino vertex
and looking for a decay kink topology in the vertex region.

\section{THE CNGS BEAM STARTUP}
 The CNGS neutrino beam~\cite{CNGS} is a high energy $\nu_\mu$ beam optimised
to maximise the $\nu_\tau$ charged current interactions at Gran Sasso produced
by oscillation mechanism at the atmospheric $\Delta m^{2}$.
The mean neutrino energy is about 17 GeV with a  contamination of 2.4\% $\overline{\nu}_{\mu}$,
 0.9\% $\nu_{e}$ and less than 0.06\% of $\overline{\nu}_{e}$.
Using the CERN  SPS accelerator in a shared mode with fixed target experiment together with LHC, 
4.5x$10^{19}$ protons on target (pot) per year should normally be delivered,
 assuming 200 days of operation.
The number of charged current and neutral current interactions expected in the Gran Sasso
laboratory from $\nu_\mu$
are then about 2900 /kton/year and 875 /kton/year respectively.
If the $\numutau$ oscillation hypothesis is confirmed, 
the number of $\tau$'s produced via
charged current interaction at the Gran Sasso should be
of the order of 14 /kton/year 
for  $\Delta m^{2}=$2.5x$10^{-3}$$ \mathrm{eV}^{2}$ at full mixing. 

A first CNGS short run took place in August 2006. The OPERA target was empty at that time
but the electronic detectors were taking data.
During this run, 319 events correlated in time with the beam and coming from neutrino 
interactions in the surrounding rock and inside the detector have been recorded. 
 The delivered intensity correponded to  $7.6\mathrm{x}10^{17}$ pot, with a 
peak intensity of $1.7$x$10^{13}$ pot per extraction corresponding to 70\% of the expected nominal value.
 The reconstructed zenith angle distribution from penetrating muon tracks was showing a clear
peak centered around $3.4^{o}$ as expected for neutrinos originating from CERN.
Details and results can  be found in Ref~\cite{cngs2006}.

\section{FIRST NEUTRINO EVENTS AND DETECTOR PERFORMANCES}
\begin{figure*}[htb]
\vspace{-0.3cm}
\begin{center}
    \includegraphics[width=10cm]{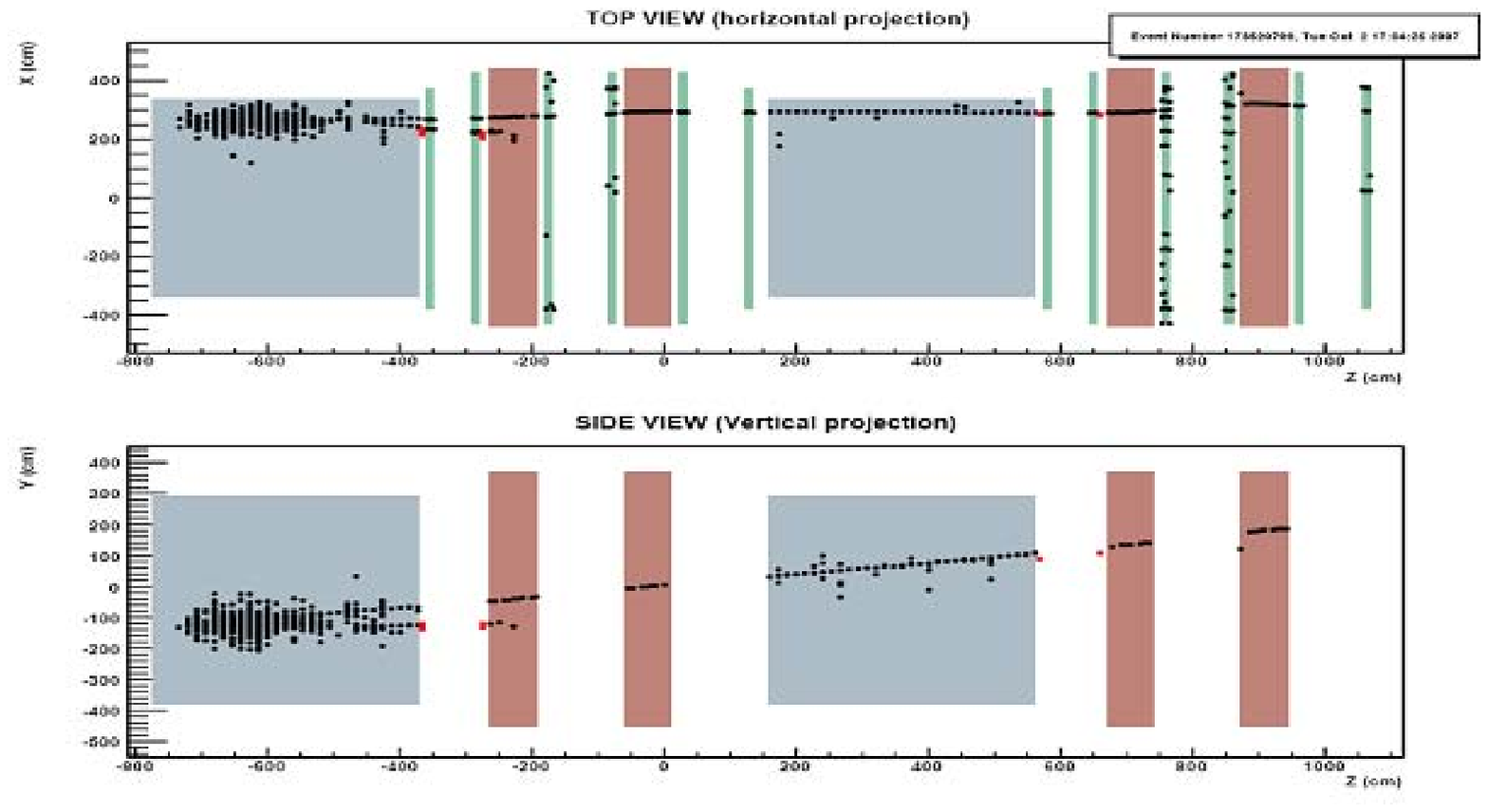}
    \includegraphics[width=7cm]{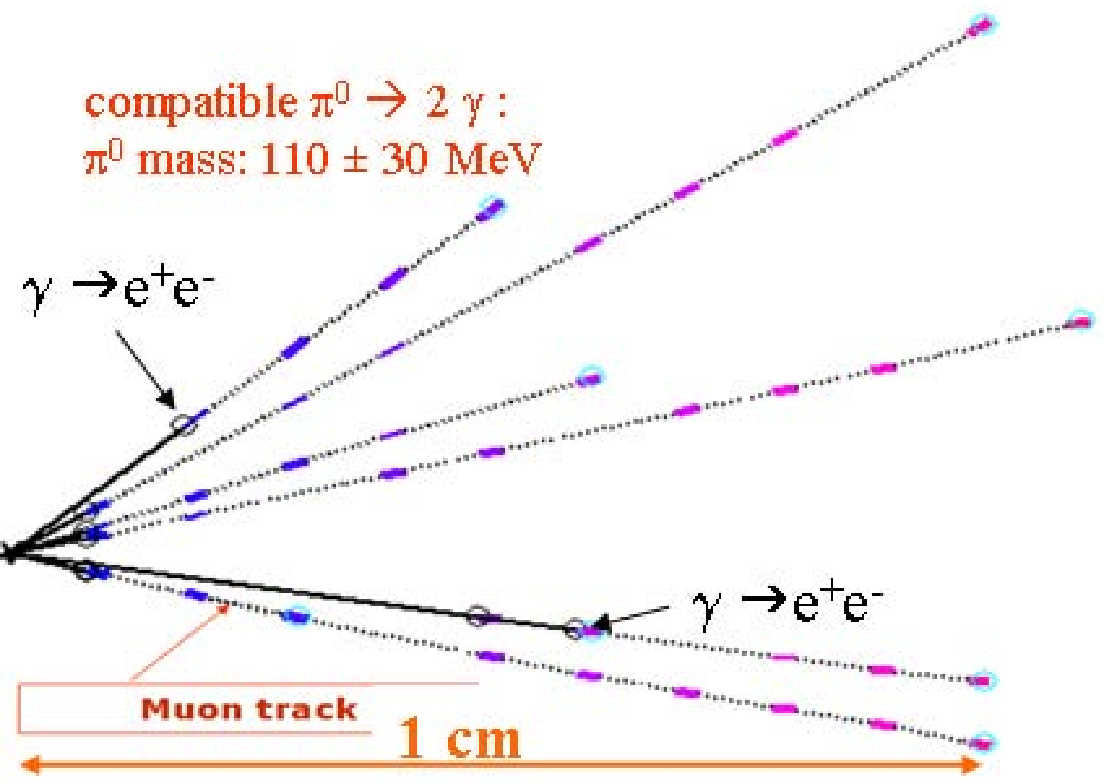}

\end{center}
\vspace{-0.8cm}
\caption{a) Charged current neutrino interaction recorded in OPERA. The event display 
shows the hits left in the electronic detectors). b) Emulsion reconstruction of the 
neutrino interaction vertex in the corresponding target brick.}
\label{fig:cngs_event}
\vspace{0cm}
\end{figure*}

A second CNGS physics run took place in October
2007 with a total of  $8.24\mathrm{x}10^{17}$ pot delivered and 369 reconstructed beam related events.
Similar selection criteria to the 2006 analysis~\cite{cngs2006},
 based on GPS timing systems and synchronisation between OPERA and CNGS, have been used
to select events compatible with the CNGS proton extraction time window. 
 The OPERA target was filled with 80\% of the first
supermodule corresponding to a total target mass of 0.5 kton. 
Among the selected beam events, 38 were recorded and reconstructed inside the OPERA target
for 31.5$\pm$6 expected.
Among them, 29 were classified as Charged Current (CC) and 9
as Neutral Current (NC) in agreement with expectation. 
For each event the electronic detector hits were used to find the most probable
brick where the neutrino interaction may have occured. 
The left part of Figure~\ref{fig:cngs_event} shows an event display of the first neutrino 
interaction located in the OPERA detector. The black dots represent hits in the 
electronic detector. The event is a charged current event with a clear muon track traversing
both target and spectrometer sections over more than 18 m. The right part of the figure
shows the result of the detailed analysis of the emulsions after scanning the identified 
brick where a clear reconstructed
interaction vertex is visible with two photon conversions compatible with 
a $\pi^{o}$ decay.

\begin{figure*}[htb]
\vspace{-0.3cm}
\begin{center}
    \includegraphics[width=9cm]{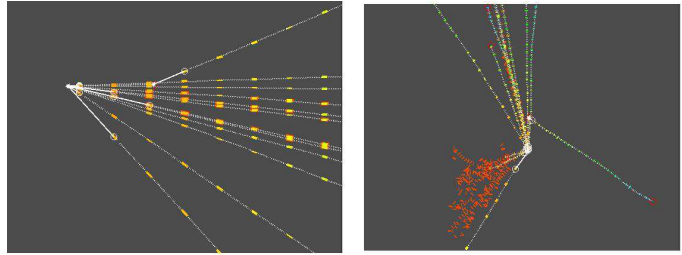}
\end{center}
\vspace{-0.8cm}
\caption{Longitudinal and transverse view of a reconstructed neutrino interaction vertex with a charm
decay candidate topology.}
\label{fig:charm2007}
\vspace{0cm}
\end{figure*}

The extensive study of the recorded events have confirmed the OPERA
performances and the validity of the methods and algorithms used which, for example, give
impact parameter resolution of the order of a few microns,  particle momentum estimation, 
shower detection for e/$\pi$ separation.
 Figure~\ref{fig:charm2007} shows the longitudinal and 
transverse views of another reconstructed event vertex where a clear decay topology similar to what is expected
from a $\tau$ decay is visible. However, the presence of a prompt muon attached to the primary vertex and the momentum
balance in the transverse plane is in favour of a $\nu_\mu^{CC}$ interaction producing a charm particle. 
 
\section{CONCLUSIONS}

The OPERA detector is completed and is now massive with 1.25 kton of lead-emulsion target offering
a huge and precise tracking device. 
With the cosmic data taking and the first CNGS neutrino runs in 2006 and 2007,
the design goals and detector perfomances
were reached and the first levels of the reconstruction software and analysis tools were validated.
The observation in 2007 of 38 neutrino events in the target bricks, the 
localization and reconstruction of neutrino vertex in emulsions was an important phase which
succesfully validated the OPERA detector concept.\\
Having now the full OPERA target, the next important step is the 2008 CNGS neutrino run which
started already  in June.
It is expected to have about $2.28$x$10^{19}$ pot in 123 days of SPS running assuming a nominal 
intensity of $2$x$10^{13}$ pot/extraction. This intensity, when reached, should lead
 to about 20 neutrino interactions/day  in the target and eventually the  
observation of the first $\tau$ event candidate.\\
In 5 years of CNGS running at 4.5x$10^{19}$ pot per year, OPERA should be able 
to observe 10 to 15 $\nu_\tau$ events after oscillation at
full mixing in the range $2.5$x$10^{-3} < \Delta m^{2} <$ 3x$10^{-3}$ $\mathrm{eV}^{2}$,
with a total background less than 0.76 events.

\end{document}